\documentclass[prl,twocolumn,superscriptaddress,preprintnumbers,amsmath,amssymb]{revtex4}
\usepackage[dvips]{graphicx}
\usepackage{dcolumn}
\usepackage{color}
\usepackage{bm}

\usepackage{epstopdf}

\begin{document}

\title{Coulomb gap refrigerator}

\author{Jukka P. Pekola}
\affiliation{Low Temperature Laboratory (OVLL), Aalto University School of Science, P.O. Box 13500, 00076 Aalto, Finland}
\author{Jonne V. Koski}
\affiliation{Low Temperature Laboratory (OVLL), Aalto University School of Science, P.O. Box 13500, 00076 Aalto, Finland}
\author{Dmitri V. Averin}
\affiliation{Department of Physics and Astronomy, Stony Brook University, SUNY, Stony Brook, NY 11794-3800,
USA}

\begin{abstract}
We propose a remarkably simple electronic refrigerator based on the Coulomb barrier for single-electron tunneling. A fully normal single-electron transistor is voltage $V$ biased at a gate position such that tunneling through one of the junctions costs an energy of about $k_BT \ll eV, E_C$, where $T$ is the temperature and $E_C$ is the transistor charging energy. The tunneling in the junction with positive energy cost cools both the electrodes attached to it. Immediate practical realizations of such a refrigerator make use of Andreev mirrors which suppress heat current while maintaining full electric contact.
\end{abstract}

\date{\today}

\maketitle
Thermal transport properties of nanocircuits are receiving increased attention \cite{giazotto06,muhonen12}. Overheating due to dissipative currents is a concern for applications with either dense architecture or when operating in a regime where thermal relaxation becomes weak, for instance at low temperatures. Active cooling below the bath temperature is one of the available strategies against overheating, and can be achieved directly by electric means on a chip. The practical realizations employ energy-selective transport either with the help of a superconducting gap \cite{nahum94,leivo96,ullom13,giazotto06} or via a discrete level in a quantum dot \cite{edwards93,prance09,gasparinetti11}. Here we present a basic, till now overlooked alternative method based on the mere Coulomb gap in a simple single-electron transistor with metallic electrodes \cite{averin86,set}. The overall dissipation of a biased normal single-electron transistor is naturally positive, but due to the Coulomb gap we can find regimes where one of the junctions cools the lead and the island whereas the other one is dissipative. This provides an interesting possibility for realizing a Coulomb blockade enabled refrigerator ("SET cooler"), if the charge and energy degrees of the single-electron transistor can be controlled independently, e.g., if the transistor island can be split by a superconducting inclusion in two halves thermally while maintaining its electric unity. Although operation of such a cooler is based on electrostatic energy gap for electron transport similarly to the superconducting gap and quantum dot coolers, which also use the energy gaps, the nature of the electrostatic gap makes the SET cooler different from them in one important respect. While those coolers can be viewed in some respects as Peltier-effect refrigerators (see, e.g., \cite{thel}) in which only one electrode of the tunnel junction is cooling down while the other one is heating, the removed heat in the SET cooler is split equally between the two electrodes of the cooling junction. An attractive feature of the SET cooler is the possibility to adjust the gap by gate voltage to optimize the operation at a given temperature. We discuss the performance of the refrigerator in detail and potential ways to realize it in practice. It turns out that the SET cooler is most suitable for very low temperatures, where the standard superconducting gap based electronic coolers become inefficient \cite{giazotto06,muhonen12}.

\begin{figure}
    \begin{center}
    \includegraphics[scale=.45]{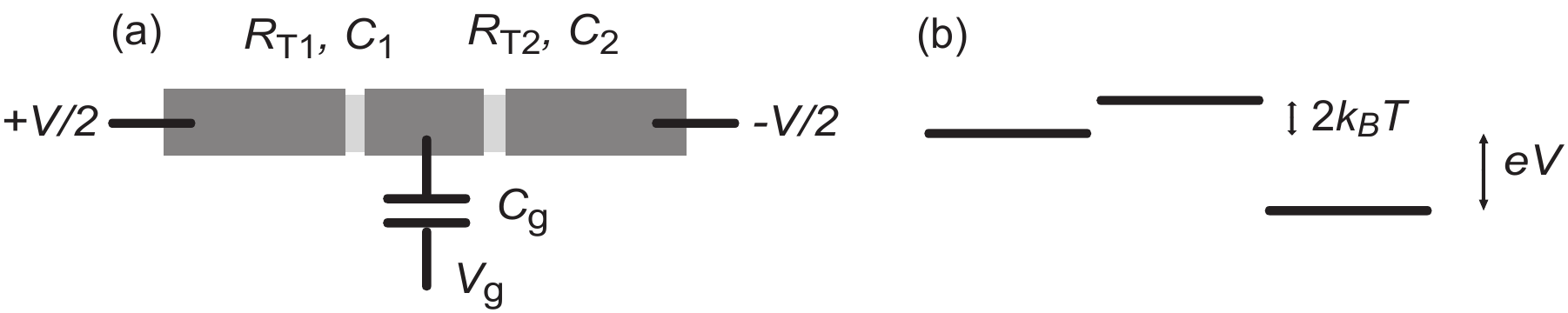}
    \end{center}
    \caption{The single-electron transistor as a cooler. In (a) the basic structure is shown with bias voltage $V$, tunnel junctions with capacitances $C_i$ and resistances $R_{Ti}$, and gate at voltage $V_g$ and capacitance $C_g$. In the text we mostly assume the structure to be symmetric with $R_{Ti} = R_T$ and $C_i=C$ for both junctions. In (b) we demonstrate the biasing for optimum cooler operation in the two-state approximation, where the energy cost to tunnel through the first (cooling) junction is $2k_BT$, and $-(eV+2k_BT)$ through the second one.}
    \label{fig:SETcooler}
\end{figure}

Figure \ref{fig:SETcooler} shows the basic scheme, where a standard single electron transistor is biased at voltage $V$, and its gate position is $n_g \equiv -C_gV_g/e$, where $C_g$ and $V_g$ are the gate capacitance and voltage, respectively. We analyze the energetics of the single-electron transistor, giving basic analytic results in the low temperature regime  $k_BT \ll E_C$, where only two charge states $n=0$ and $n=1$, are possible.  For optimal operation in this regime, the gate voltage is adjusted to a value where the in-tunneling electron experiences a barrier $\sim k_BT \ll eV$, where $T$ is the temperature of the electrodes, and the out-tunneling electron experiences an energy gain $\sim eV$. Under these conditions the electrodes attached to the junction of the former tunneling event experience cooling and those to the latter one heat up. We consider arbitrary gate positions within $0 < n_g < 1$. Due to simple symmetries, the roles of the two junctions are interchanged when operating the single-electron transistor at the gate position $1-n_g$ instead of $n_g$.

We write first the equations governing the charge and energy dynamics of the single-electron transistor, but here limiting to equal temperatures in all electrodes. The rates of single-electron tunneling into ($+$) or out ($-$) of the island through junction $k=1$ or $k=2$ in the charge state $n$ are given by
$\Gamma_k^\pm (n) = (e^2R_T)^{-1}\Delta E_k^\pm(n)/(e^{\beta\Delta E_k^\pm(n)}-1)$,
where $R_T$ is the tunnel resistance of the junctions that is for the moment assumed to be the same for the two junctions, $R_{T1}=R_{T2}=R_T$,  and
$\Delta E_k^\pm(n)=\pm (-1)^keV/2\pm 2E_C(n-n_g\pm 1/2)$
are the energy costs for the various processes. $E_C=e^2/2C_\Sigma$ is the magnitude of the charging energy, and the common temperature is given by $T=(k_B\beta)^{-1}$. Here, $C_\Sigma = 2C + C_g$ is the total capacitance of the structure, and $C$ is the capacitance of one junction (again assuming a symmetric structure).
The corresponding occupation probabilities $p(n)$ obey the steady-state result
$[\Gamma_1^+(n-1)+\Gamma_2^+(n-1)]p(n-1)=[\Gamma_1^-(n)+\Gamma_2^-(n)]p(n)$
normalized by $\sum_{-\infty}^\infty p(n)=1$.

The heat currents corresponding to the various processes can be written as
\begin{equation} \label{heat1}
\dot Q_k^\pm(n)= \mp \frac{1}{e^2R_T}\int dE \, E\,f_{L,k}(\pm E+\Delta E_k^\pm(n))[1-f_I(\pm E)]
\end{equation}
for the partial cooling power by the tunneling into ($+$) and out from ($-$) the island.
Here, $f_{I/L,k}(E)$ are the energy distributions (typically Fermi distributions) of the island $I$ and the leads $L,k$, respectively. The total heat current out from the island (= cooling power) through each junction is then given by
$\dot Q_k = \sum_{n=-\infty}^\infty p(n)[\dot Q_k^-(n)+\dot Q_k^+(n)].$
This is also the cooling power for the corresponding lead attached to junction $k$: as one can see from Eq. \eqref{heat1}, the heat extracted from or released into the junction electrodes in a tunneling process is the same for both electrodes of the junction. For Fermi distributions, assuming again all temperatures to be the same, Eq. \eqref{heat1} can be integrated analytically with the result
\begin{equation} \label{heat3}
\dot Q_k^\pm(n)= \frac{1}{2 e^2R_T}\frac{[\Delta E_k^\pm(n)]^2}{e^{\beta \Delta E_k^\pm (n)}-1}.
\end{equation}

Next we focus on the two state regime at low temperatures, $k_BT\ll E_C$ in the gate interval $0<n_g<1$. Furthermore, we assume that the bias voltage is large enough, $eV\gg k_BT$, such that all tunneling occurs in the "forward" direction. Then we need to consider only two single-electron processes, $+$ for $n=0 \rightarrow 1$ and $-$ for $n=1 \rightarrow 0$ transition, respectively, with energy costs
$\Delta E^\pm=-\frac{eV}{2}\pm 2E_C(\frac{1}{2}-n_g)$,
and occupations
$p(1)=1-p(0)= \Gamma^+/(\Gamma^++\Gamma^-)$.
Within this two-state approximation, we notice that based on Eq. \eqref{heat3}, remembering that $k_BT\ll E_C$, the cooling power of the first junction obtains the maximum value when the barrier has the magnitude $\Delta E^+ \simeq 2k_BT$ indicated in Fig. \ref{fig:SETcooler}. The cooling power of one side of the optimally biased junction is then given approximately by
\begin{equation} \label{powera}
\dot Q_{\rm opt} \simeq 0.31 \frac{(k_BT)^2}{e^2R_T},
\end{equation}
The gate position for maximum cooling is given by
\begin{equation} \label{opt}
n_g^{\rm opt} -1/2 = \mp (\frac{k_BT}{E_C}+ \frac {1}{4}\frac{eV}{E_C}).
\end{equation}
\begin{figure}
    \begin{center}
    \includegraphics[scale=.33]{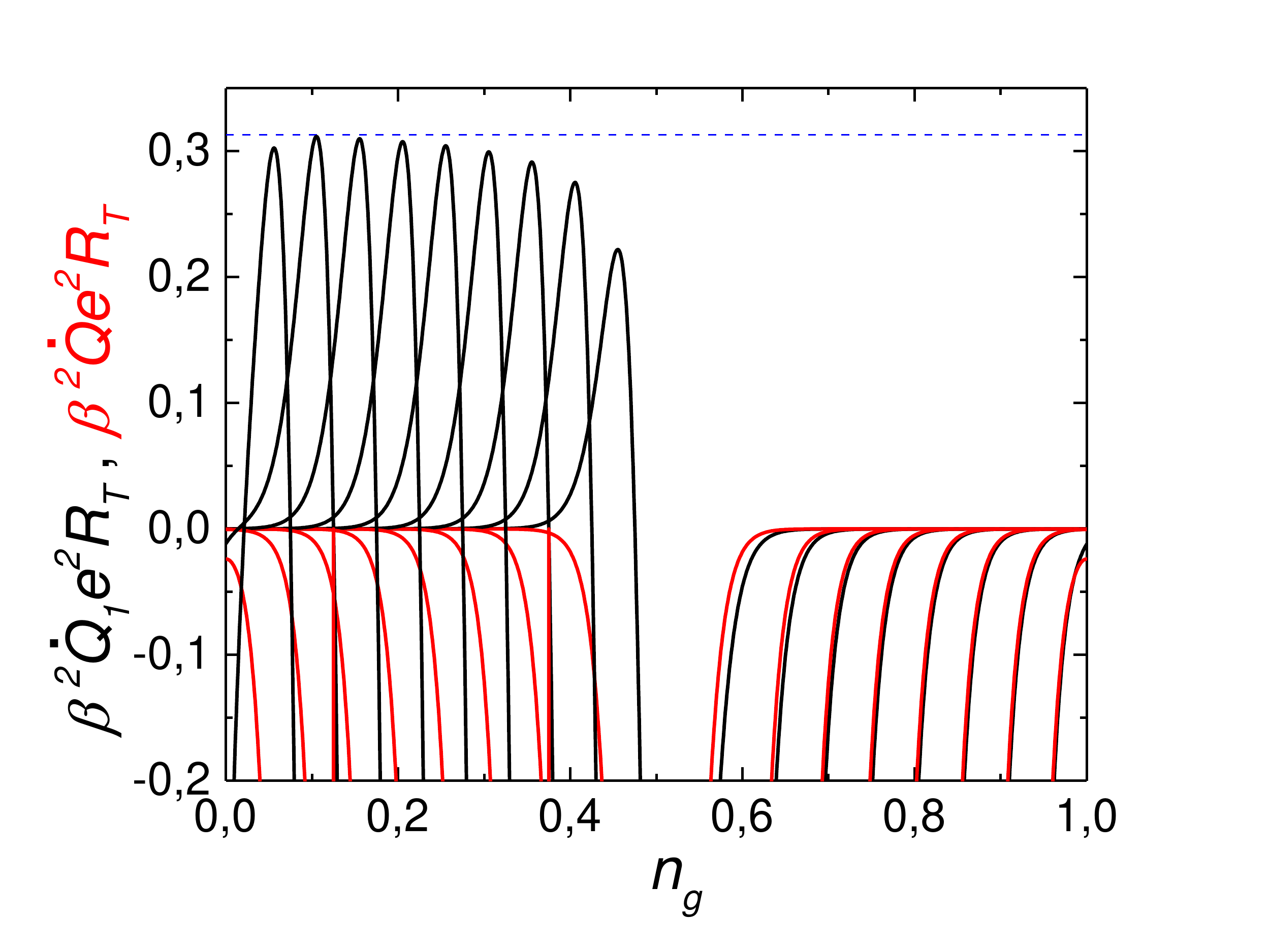}
    \end{center}
    \caption{The normalized cooling power of each side of junction $1$, $\dot Q_1/[(k_BT)^2/(e^2R_T)]$ (black), and the normalized total cooling power on the island by the two junctions (red), $\dot Q/[(k_BT)^2/(e^2R_T)]$. The latter quantity is naturally always negative with the value $\dot Q= -IV/2$. The parameters of the system are $k_BT/E_C = 0.025$, $eV/E_C = 0.1,0.3,0.5,0.7,0.9,1.1,1.3,1.5,1.7$ for curves with cooling maxima in the black curves shifting from right to left. Above this value of voltage, the achievable cooling power diminishes quickly. The dashed horizontal line is the analytic prediction of the optimum cooling power, Eq. \eqref{powera}.}
    \label{fig:dotQ}
\end{figure}
Further in the two-state approximation, the total dissipation in the biased device, $P\equiv -2\dot Q=-2\dot Q^+-2\dot Q^-$ equals $IV$, independent of the gate position. In an arbitrary position within the given gate interval the cooling power of each side of junction $1$ is given by
\begin{eqnarray} \label{j1}
&& \dot Q_{1} = p(0)\dot Q^+ = \frac{1}{2}\frac{I}{e} \Delta E^+.
\end{eqnarray}
Here $I = e \Gamma^+\Gamma^-/[\Gamma^++\Gamma^-]$ is the current through the single-electron transistor.

Within the same approximation the efficiency of the cooler obtains a natural value
\begin{eqnarray} \label{efficiency}
&& \eta= \frac{\dot Q_1}{IV} = \frac{1}{2} \frac{\Delta E^+}{eV}
\end{eqnarray}
for one side of the cooling junction, and twice this value for the entire cooling junction.
At the optimum working point of Eq. \eqref{opt}, we have $\eta_{\rm opt}= k_BT/eV$.

Pure numerical evaluation of the equations above in the general situation, not limited to two charge states only, is straightforward. The resulting cooling powers, still assuming equal temperature to all the electrodes, are given in Fig. \ref{fig:dotQ} for a realistic set of parameters. The optimum cooling power in the two-state model, Eq. \eqref{powera}, is shown by the dashed line, and it compares favourably with the numerically obtained peak cooling power.

We next consider the influence of the higher-order processes on cooling in this device. The heat current by cotunneling can be obtained by appropriately adapting the corresponding rates of charge transport \cite{averin90,set}. We may thus write the "cooling power" by cotunneling for instance in the electrodes $\ell=1,2$ attached to the first junction [$\ell=1$ for the external lead connected to junction $1$, $\ell =2$ for the electrode connected to junction $1$ on the island] as
\begin{widetext}
\begin{eqnarray} \label{cotunneling1}
&& \dot Q_\ell^{\rm c.t.}(n) = \frac{1}{2\pi \hbar}(\frac{R_Q}{R_T})^2\int d\epsilon_1 d\epsilon_2 d\epsilon_3 d\epsilon_4 (-1)^{\ell -1} \epsilon_\ell f(\epsilon_1)[1-f(\epsilon_2)]f(\epsilon_3)[1-f(\epsilon_4)]\times \nonumber\\&&
\Big(\frac{1}{\epsilon_2-\epsilon_1 + \Delta E_1^+(n)}+\frac{1}{\epsilon_4-\epsilon_3+\Delta E_2^-(n)}\Big)^2\delta(eV+\epsilon_1-\epsilon_2+\epsilon_3-\epsilon_4).
\end{eqnarray}
Here $R_Q =\hbar/e^2\simeq 4.1$ k$\Omega$. The integrals over three of the four energies can be done analytically, and the remaining one reads
\begin{eqnarray} \label{cotunneling2}
&& \dot Q_\ell^{\rm c.t.}(n) = - \frac{1}{4\pi \hbar}(\frac{R_Q}{R_T})^2\int d\epsilon \frac{eV+\epsilon}{1-e^{-\beta (eV+\epsilon)}}\frac{\epsilon^2}{1-e^{\beta \epsilon}}\Big(\frac{1}{\epsilon - \Delta E_1^+(n)}-\frac{1}{\epsilon+\Delta E_2^-(n)+eV}\Big)^2
\end{eqnarray}
for both $\ell =1$ and $\ell=2$.
\end{widetext}
At zero temperature, for the Coulomb blockade conditions, $eV \ll \Delta E_1^+ (0), \Delta E_2^-(0)$, we obtain
\begin{eqnarray} \label{cotunneling3}
&& \dot Q_\ell^{\rm c.t.} = -\frac{1}{48\pi \hbar}(\frac{R_Q}{R_T})^2\Big(\frac{1}{E_1}+\frac{1}{E_2}\Big)^2 (eV)^4.
\end{eqnarray}
Here, $E_1 \equiv \Delta E_1^+(0)$ and $E_2 \equiv \Delta E_2^-(0)$ if the single-electron tunneling is pinched-off by the first junction ($n=0$ state dominates). The quantity in Eq.~\eqref{cotunneling3} is always negative, i.e cotunneling results in heating, which is naturally small for $R_T \gg R_Q$ and small values of $V$.

Next we analyze the most relevant regime close to the optimum cooling bias for junction 1, when $\beta E_1\sim 2$, while $E_2 \sim E_C\gg \beta^{-1}, E_1$, and $eV\gg \beta^{-1}, E_1$. Then the term with $E_1$ in the denominator dominates Eq.~\eqref{cotunneling3}, i.e., the cotunneling goes predominantly through one intermediate charge state ($E_1$), making it possible to simplify the equations by neglecting the processes through the other charge state, with energy $E_2$. On the other hand, description of cotunneling in the regime with $\beta E_1\sim 2$ is complicated by the fact that for such a small charging energy barrier, sequential "first-order" classical tunneling {\em over} the barrier cannot be clearly separated from the cotunneling, which is the "second-order" tunneling {\em through} the barrier (cf. Fig.~1). In general, coexistence of the tunneling events of different order requires taking into account the non-perturbative effect of broadening of the charge states by tunneling \cite{cbt}. In the situation of the optimum cooling bias, $eV\gg \beta^{-1}, E_1$, the broadening of the relevant charge state $E_1$ is dominated by tunneling in the second junction.

Quantitatively, employing the usual tunnel Hamiltonian $H_T$, we can express the average of the cooling power $\dot Q_1$ as
\begin{equation}
\langle \dot Q_1 \rangle = \langle U^{\dagger}(t) \dot Q_1 (t) U (t) \rangle\, , \; U (t) =\mathcal T\exp \{\frac{-i}{\hbar}\int^{t} dt' H_T (t')\} \, .
\label{a1} \end{equation}
Here the time dependence of all operators is due to the charging energy of the transistor and internal energy of the electrodes, the average $\langle ... \rangle$ is taken over the assumed equilibrium state of the electrodes, $\mathcal T$ denotes time-ordering, and, in the standard notations,
\[ \dot Q_1 = \frac{i}{2\hbar} \sum_{k,p} (\epsilon_k - \epsilon_p) [t_{kp}^{(1)} c_k^{\dagger} c_p- h.c.] \, , \;\; H_T=H_1+H_2 \, , \]
where $H_2= \sum_{q,l}[t_{q,l}^{(2)} c_q^{\dagger} c_l+ h.c.]$, and a similar expression for the tunneling Hamiltonian $H_1$ of the first junction.

In the regime described qualitatively above, Eq.~(\ref{a1}) can be evaluated expanding the evolution operator $U(t)$ to the lowest power in $H_1$, but summing the main terms that correspond to broadening of $E_1$ to all powers in $H_2$. (In this calculation, we allow the two junction conductances $G_{1,2}=1/R_{T1,2}$ to be in general different.) This gives, dropping $\langle...\rangle$ out for simplicity in notation,
\[ \dot Q_1 =\frac{G_1}{2\pi e^2}\int d\epsilon \frac{\epsilon^2}{1-e^{-\beta\epsilon}} \Im{\rm m} (\Sigma_{n=0}^{\infty} \frac{\xi^n}{(\epsilon+E_1+i0)^{n+1}})\, , \]
where
\[ \xi= \frac{\hbar G_2}{2\pi e^2}\int d\epsilon' \frac{\epsilon'}{1 -e^{-\beta\epsilon'}}\frac{1}{\epsilon+\epsilon'-eV+i0} \, . \]
The real part of $\xi$ contributes to the tunneling-induced shift of the energy of the intermediate charge state. Incorporating it into the actual energy of this state: $E_1 \rightarrow E$, one is left with the broadening of this state by the imaginary part of $\xi$
\begin{equation}
\Im {\rm m}\, \xi= \frac{\hbar G_2}{2e^2} \frac{eV-\epsilon}{1- e^{-\beta(eV- \epsilon)}} \equiv \gamma (\epsilon)\, , \label{a2} \end{equation}
i.e., the level is broadened to the width $\gamma$ which coincides
with the half of the tunneling rate in the second junction at bias
$eV- \epsilon$. Taking into account that $\beta eV\gg 1$, one obtains then the following final expression for the cooling power:
\begin{equation}
\dot Q_1 = \frac{\hbar G_1 G_2}{4 \pi e^4}\int_{-\infty}^{eV} d\epsilon \frac{\epsilon^2}{1-e^{-\beta \epsilon}} \frac{\epsilon - eV}{(\epsilon+E)^2 +\gamma^2(\epsilon)} \, .
\label{a3} \end{equation}

\begin{figure}
    \begin{center}
    \includegraphics[scale=.35]{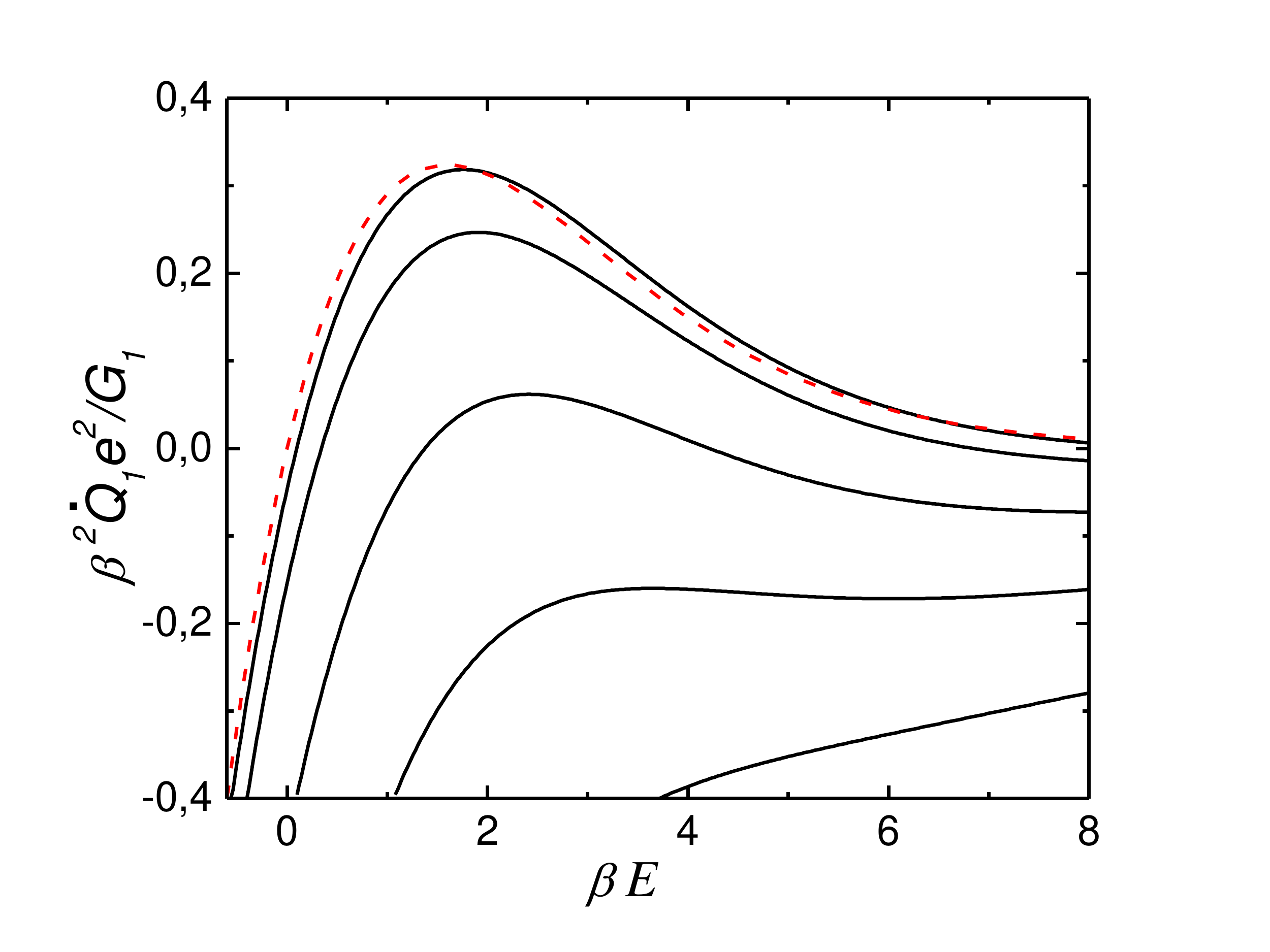}
    \end{center}
    \caption{The total power $\dot Q_1$ as a function of the charging energy barrier $E$ for tunneling in this junction, for several values of $\gamma$, the energy width of the intermediate charge state created by tunneling in the second junction. From top to bottom, $\beta\gamma = 0.03; 0.1; 0.3; 0.6; 1.0$. For all curves, the bias voltage is $\beta eV = 20$. The dashed red line is the ideal classical cooling power. The plot illustrates increasing cotunneling contribution to heating in junction $1$ and the simultaneous broadening of the classical cooling peak with increased width $\gamma$. }
    \label{fig:P}
\end{figure}

In the limit of interest, $eV\gg \gamma, \beta^{-1}, E$, the energy dependence of $\gamma$ can be neglected, $\gamma=\gamma(\epsilon=0)=\hbar G_2V/2e$, and the integral in Eq.~(\ref{a3}) can be evaluated in terms of the digamma function $\psi (z)$ as
\begin{eqnarray} \label{a4}
&&\dot Q_1 = \frac{\hbar G_1 G_2}{2\pi e^4} [eV E (\ln \frac{\beta eV}{2\pi} -1)-\frac{(eV)^2}{4}] + \\&&
\frac{G_1}{4 e^2} \{\frac{2}{\pi} \Im {\rm m} [ (E-i\gamma)^2\psi(\frac{\beta(E+i\gamma)}{2\pi})]-E^2-2\beta^{-1}E+\gamma^2 \} \nonumber .
 \end{eqnarray}
This result is plotted in Fig.~\ref{fig:P}. For $\gamma\rightarrow 0$,  Eq.~(\ref{a4}) reproduces the classical result of Eq. \eqref{heat3}, $\dot Q_1=(G_1E^2/2e^2)/(e^{\beta E}-1)$ which also closely approximates the top numerical curve in Fig.~\ref{fig:P}. We see, both from Eq.~(\ref{a4}) and Fig.~\ref{fig:P} that the effect of the higher-order tunneling processes on cooling includes direct cotunneling-induced heating (the first line in Eq.~(\ref{a4})) and broadening and suppression of the classical cooling peak by the level width $\gamma$ (the second line). Direct cotunneling heating is small as long as $\gamma \ll (eV\beta^2)^{-1}$, while the broadening is almost negligible for $\beta\gamma<0.1$. Elsewhere in this paper  we assume that these conditions are satisfied and we can use the classical description of cooling.

\begin{figure}
    \begin{center}
    \includegraphics[scale=.35]{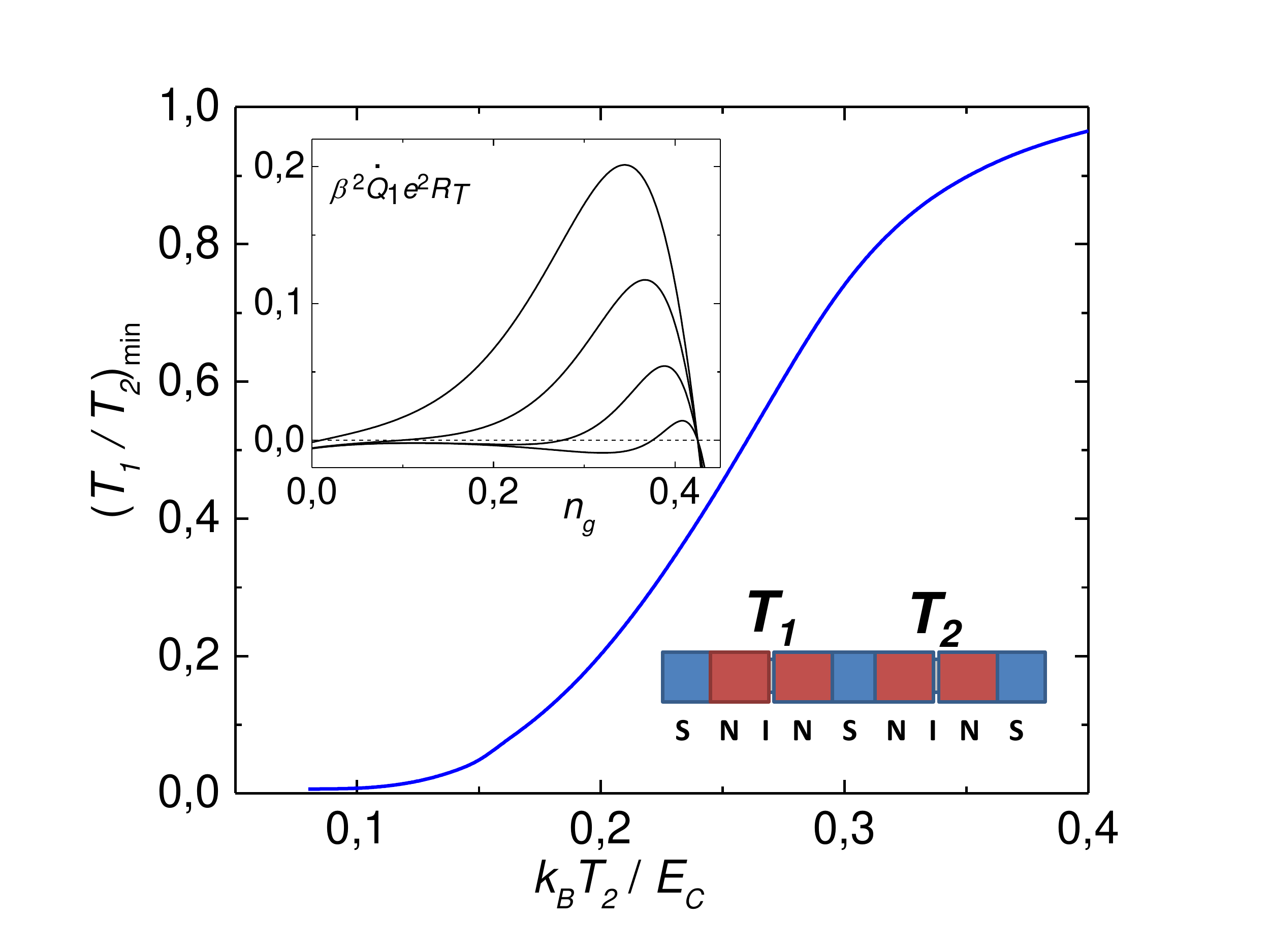}
    \end{center}
    \caption{Cooling under the conditions of unequal temperatures. The lower inset shows the sketch of a SET cooler with superconducting (S) mirrors separating normal (N) conductors. The temperatures around junctions $1$ and $2$ are given by $T_1$ and $T_2$, respectively, and they are assumed to be the same on both sides of each junction (I). The upper inset shows the gate voltage dependence of the cooling power for $eV=0.3E_C$, $k_BT_2/E_C=0.1$, and $k_BT_1/E_C = 0.1,0.075,0.05,0.025$ from top to bottom. The main frame shows the temperature drop $(T_1/T_2)_{\rm min}$ for zero cooling power for $eV=0.3E_C$ as a function of $T_2$.}
    \label{Fig4}
\end{figure}

Next we turn to the practical realization of the cooler. In general, the cooling effect (temperature drop) is unnoticeable in a standard single-electron transistor, because the lead electrodes are reservoirs thermalized by large volume and by effective heat conduction near the junction, and, on the other hand, the total power on the island is positive. However, it is quite straightforward to realize a configuration, where the charge and heat currents separate effectively. The most obvious way to do this is to replace parts of the normal electrodes by superconductors (forming Andreev mirrors with direct metal-to-metal contacts) that efficiently isolate the cooled areas without influencing the relevant charge transport in the cooler \cite{nahumandreev,golubev01,peltonen10}. This can be done by splitting the island into two halves, and by interrupting one or both the leads this way, see the lower inset in Fig. \ref{Fig4}. In this configuration it is more practical to cool and monitor the normal section of the lead outside the transistor island. This makes the thermometry, e.g. by tunnel spectroscopy, and other measurements straightforward, because then the potential of the cooled area does not vary in response to individual tunneling events.

The fundamental limitation of the performance of the SET cooler in terms of the minimum temperature is given by the temperature $T_2$ of the "hot" junction. The cooling of junction $1$ (at temperature $T_1$) diminishes, as more charge states become available due to tunneling in the higher temperature junction, and eventually there will be power $IV/4$ deposited to all the four electrodes when the Coulomb effects become negligible. Naturally this is not the only limitation on cooling, other mechanisms include heat load from the phonon bath and through the superconducting lead to the cooled area, but the latter contributions can be made small by operating at low temperatures and by proper choice of the geometries of the device. The second inset in Fig. \ref{Fig4} shows as an example a set of cooling powers of junction $1$ at various values of $T_1 \ll T_2$, plotted again as a function of gate voltage at a fixed bias voltage $V$. Naturally the power gets smaller on reducing $T_1$ because of the backflow of heat from the hot bath, and since the (cooling) current of the device decreases on decreasing $T_1$. The main frame of Fig. \ref{Fig4} shows the ultimate achievable temperature reduction $(T_1/T_2)_{\rm min}$ as a function of $T_2$, given by the minimum value of $T_1$ where the cooling power gets positive values at the optimum point. We see that temperature reductions by an order of magnitude seem feasible from this point of view.

Finally we give a few practical remarks. It is favourable to increase the value of $E_C$ as high as is practical in order to keep the device in the SET regime with just two charge states. With the conventional metallic realization of the circuit, values of $E_C/k_B \sim 1-3$ K can be achieved in a single-electron transistor whose island is several $\mu$m long. This, in turn, allows for the insertion of the superconducting mirror and sufficient volume near junction $2$ on the island to avoid excessive overheating. To make these arguments more concrete, we consider the various heat currents briefly. When a superconducting Al wire is longer than $\sim 1$ $\mu$m, the adjacent island is better coupled to the phonon bath than through the wire electronically at operating temperatures $\sim 100 $ mK, as was demonstrated in Ref. \cite{peltonen10}. Thus the cooling properties are not much affected by the heat leak through the Al wire. We equate the ideal cooling power \eqref{powera}, and the standard heat load $\Sigma \mathcal V (T_p^5-T^5)$ from the phonons, where $T_P$ is the temperature of the phonon bath, $\Sigma = 2\times 10^{9}$ WK$^{-5}$m$^{-3}$ for copper as the normal metal \cite{giazotto06}, and $\mathcal V = 10^{-21}$ m$^3$ is the volume of the cooled electrode. With these parameters, it should be possible to reach $T_1$ as low as 10 mK with $R_T=1$ M$\Omega$ at the bath temperature of $T_p=50$ mK. On the other hand, the island near junction $2$ would warm up to a temperature $T_2 \simeq [P/(\Sigma \mathcal V_2)]^{1/5}$, where $P\simeq IV/2$ is the Joule power due to dissipative tunneling in junction $2$ and $\mathcal V_2$ is the volume of the normal island near this junction. We obtain $T_2\sim 100$ mK, still compatible with $T_1=10$ mK based on Fig. \ref{Fig4}. The cotunneling heating is low when $eV$ is chosen properly (at such low temperatures the $V$ dependence of cooling is weak even below $eV=0.1E_C$).

In summary, we have proposed and analyzed a new low temperature electronic cooler based on an adjustable Coulomb gap in a single-electron transistor.

The work has been supported partially by the Academy of Finland
through its LTQ CoE grant (project no. 250280), and the European
Union FP7 project INFERNOS (grant agreement 308850).


\begin{thebibliography}{99}
\bibitem{giazotto06} F. Giazotto, T. T. Heikkil\"a, A. Luukanen, A. M. Savin, and J. P. Pekola, Rev. Mod. Phys. {\bf 78}, 217 (2006).

\bibitem{muhonen12} J. T. Muhonen, M. Meschke, and J. P. Pekola, Rep. Prog. Phys. {\bf 75}, 046501 (2012).

\bibitem{nahum94} M. Nahum, T. M. Eiles, and John M. Martinis, Appl. Phys. Lett. {\bf 65}, 3123 (1994).

\bibitem{leivo96} M. M. Leivo, J. P. Pekola, and D. V. Averin, Appl. Phys. Lett. {\bf 68}, 1996 (1996).

\bibitem{ullom13} P. J. Lowell, G. C. O'Neil, J. M. Underwood, and J. N. Ullom, Appl. Phys. Lett. {\bf 102}, 082601 (2013).

\bibitem{edwards93} H. L. Edwards, Q. Niu, and A. L. de Lozanne, Appl. Phys. Lett. {\bf 63}, 1815 (1993).

\bibitem{prance09} J. R. Prance, C. G. Smith, J. P. Griffiths, S. J. Chorley, D. Anderson,
G. A. C. Jones, I. Farrer, and D. A. Ritchie, Phys. Rev. Lett. {\bf 102}, 146602 (2009).

\bibitem{gasparinetti11} S. Gasparinetti, F. Deon, G. Biasiol, L. Sorba, F. Beltram, and F. Giazotto,
Phys. Rev. B {\bf 83}, 201306 (2011).

\bibitem{averin86} D. V. Averin and K. K. Likharev, J. Low Temp. Phys.  {\bf 62}, 345 (1986).

\bibitem{set} {\it Single Charge Tunneling}, NATO Advanced Study Institute, Ser. B, Vol. 294, edited by H. Grabert and M. H. Devoret, (Plenum Press, New York, 1992).

\bibitem{thel} D.M. Rowe, ed., {\em Thermoelectrics Handbook: Macro to Nano}, (Taylor \& Francis, 2006).  

\bibitem{averin90} D.V. Averin and Y.V. Nazarov, Phys. Rev. Lett. {\bf 65}, 2446 (1990).

\bibitem{cbt} D.V. Averin, Physica B {\bf 194/196}, 979 (1994); H.
Schoeller and G. Sch\"{o}n, Phys.\ Rev. B {\bf 50}, 18436 (1994).

\bibitem{nahumandreev} M. Nahum, P. L. Richards, and C. A. Mears, IEEE Trans. Appl. Supercond. {\bf 3}, 2124 (1993).

\bibitem{golubev01} D. Golubev and L. Kuzmin, J. Appl. Phys. {\bf 89}, 6464 (2001).

\bibitem{peltonen10} J. T. Peltonen, P. Virtanen, M. Meschke, J. V. Koski, T. T. Heikkil\"a, and J. P. Pekola,
Phys. Rev. Lett. {\bf 105}, 097004 (2010).

\end{thebibliography}
\end{document}